\documentclass[12pt,tightenlines,aps,prd,nofootinbib,superscriptaddress,eqsecnum]{revtex4}

\usepackage{amsmath,amssymb}
\usepackage[mathscr]{euscript}
\usepackage{graphicx}
\usepackage[pdftex]{color}
\usepackage[sort&compress]{natbib}
\usepackage{mathbbol}  

\usepackage[colorlinks=true,linkcolor=blue,filecolor=blue,urlcolor=blue,citecolor=blue,pdftex=true,plainpages=false]{hyperref}

\usepackage{algorithm}
\usepackage{algpseudocode}

\usepackage{xcolor}
\newcommand{\Tr}{\mathbb{Tr}}

\begin{document}

\title{Noise scheduling and linear dynamics in diffusion models on Lie groups}

\author{Javad Komijani}
\email{jkomijani@ethz.ch}

\affiliation{Institute for Theoretical Physics, ETH Zurich, 8093 Zurich, Switzerland}

\date{\today}

\begin{abstract}
We investigate the role of the noise schedule in diffusion processes on Lie groups, with particular emphasis on applications to lattice gauge theory.
We show that a specific noise schedule leads to a linear decay of the expectation value of the Wilson action as a function of diffusion time.
We compare this with Euclidean diffusion models, where such behavior requires an explicitly designed drift term, while in the Lie-group setting it arises naturally.
\end{abstract}

\maketitle

\section{Introduction}

Diffusion models on Lie groups have recently been studied in the context of lattice gauge theory as a tool for sampling gauge field configurations~\cite{Zhu:2025pmw, Aarts:2026zzr, Alharazin:2026lcb, Komijani:2026lan}.
An important ingredient of these constructions is the choice of noise schedule, which controls how the system transitions from data to noise. In ref.~\cite{Komijani:2026lan}, a specific schedule is designed to produce an approximately linear evolution of the average plaquette as a function of diffusion time; see figures 4--6 there.

In this paper, we analyze how the choice of noise schedule controls the evolution of observables under diffusion. We show that a suitable choice of schedule leads to a linear time dependence of the expectation value of the Wilson gauge action. More generally, the normalization of the noise schedule can be tuned so that Wilson loops of different sizes exhibit the same type of linear decay.

We also compare this phenomenon with diffusion models in Euclidean space, where similar linear behavior requires an explicitly designed drift term. In contrast, in the Lie group setting, this structure emerges directly from the stochastic evolution.

The manuscript is organized as follows. In Section~\ref{sec:diffusion:Lie}, we review diffusion on Lie groups and derive the evolution equation for the Wilson gauge action. In Section~\ref{sec:diffusion:scalar}, we discuss the analogous construction in Euclidean diffusion models and relate it to standard variance-preserving (VP) and sub-VP formulations. Section~\ref{sec:conclusion} contains concluding remarks.

\section{Diffusion process on Lie-group}
\label{sec:diffusion:Lie}

We consider a diffusion process on a Lie group with diffusion time $t \in [0,1]$. Let $U_0$ denote the initial Lie-group elements and $U_t$ the evolved configuration. In lattice gauge theory, $U_t \equiv U_t(x,\mu)$ represents a link variable at site $x$ and direction $\mu$.
Following~\cite{Komijani:2026lan}, the forward process is written as
\begin{align}
    U_t &= K_{t,0} U_0,\\
    K_{t, t'} &= \mathcal{T}\exp\!\left(\int_{t'}^{t} \sigma(\tau) dW_\tau\right)\,.
\end{align}
Here, $\mathcal{T}$ is the time ordering, $\sigma(\tau)$ is a scalar noise schedule,
and $dW_\tau$ is a Lie-algebra–valued increment $dW_\tau = dW^a_\tau\, T^a$,
where the coefficients $dW^a_\tau$ are independent Wiener increments, and $\{T^a\}$
are generators of the corresponding Lie algebra.

To obtain a stochastic differential equation for $U_t$, we apply It\^o calculus%
\footnote{For a pedagogical introduction to It\^o calculus, see, e.g., ref.~\cite{Gardiner:1985Handbook}.}
and obtain 
\begin{align}
  dU_t
  &= \left(\sigma(t)\, dW_t - \frac{\sigma^2(t)}{2} C_F\, dt \right) U_t \, .
  \label{eq:Ito-differential}
\end{align}
Here $C_F$ denotes the quadratic Casimir in the fundamental representation.
For generators normalized as
\begin{equation}
    \Tr(T^a T^b) = -\delta^{ab},
\end{equation}
the quadratic Casimir for $\mathrm{SU}(N)$ group is
\begin{equation}
    C_F = \frac{N^2-1}{N}\,.
\end{equation}
The key feature of~\eqref{eq:Ito-differential} is the deterministic drift term proportional to $\sigma^2(t)$, which will determine the evolution of gauge-invariant observables.

We now consider the Wilson gauge action
\begin{align}
    S_\text{W}[U_t] &=
    -\frac{\beta}{2 N} \sum_{x \in \Lambda} \sum_{\mu \neq \nu}
    \Tr\left(U_t(x, \mu) U_t(x + \hat \mu, \nu) U_t^\dagger (x + \hat\nu, \mu)U_t^\dagger (x, \nu)\right)
    \,.
    \label{eq:Wilson:Plaq}
\end{align}
Here, $U_t(x,\mu)\in \mathrm{SU}(N)$ are the link variables at diffusion time $t$, $\Lambda$ denotes the set of lattice sites, and $\beta$ is the inverse gauge coupling.
We study its expectation value $s_t = \mathbb{E}\, S_{\mathrm W}[U_t]$. Because the action is linear in each link variable, only the drift term contributes in expectation. Each plaquette contains four links, so the contribution is multiplied by a factor of four,
yielding
\begin{equation}
  ds_t = -4\left(\frac{C_F\sigma^2(t)}{2} dt\right)s_t.
\end{equation}
We now choose the noise schedule
\begin{equation}
    \sigma(t) =
    \frac{\sigma_0}{\sqrt{1 - t + \varepsilon}},
    \label{eq:noise:schedule}
\end{equation}
where $\varepsilon$ is a small regulator introduced to avoid the singularity at $t=1$.
Substituting the schedule eq.~\eqref{eq:noise:schedule} and neglecting $\varepsilon$ gives
\begin{equation}
    \frac{ds_t}{dt}
    =
    -\frac{2C_F\sigma_0^2}{1-t}\, s_t
    \quad\Rightarrow\quad
    s_t = s_0 (1-t)^{2C_F\sigma_0^2}\,.
\end{equation}
A linear decay is obtained by choosing
\begin{equation}
    \sigma_0
    =
    \frac{1}{\sqrt{2C_F}},
\end{equation}
for which
\begin{equation}
    s_t = s_0 (1 - t).
\end{equation}

More generally, a Wilson loop containing $L$ unique links leads to an overall factor $L$ in the drift contribution, and linear decay is obtained by rescaling the normalization as $\sigma_0 = 2/\sqrt{L C_F}$.
Thus, the time dependence of many observables is fully controlled by the drift induced by the stochastic evolution and can be tuned through the noise schedule.

\section{Connection to diffusion models in Euclidean space}
\label{sec:diffusion:scalar}

We now relate the linear decay of Wilson loops observed in the Lie group setting to standard diffusion models in Euclidean space.
In particular, we consider variance-preserving (VP) and sub-VP formulations commonly used in score-based generative modeling~\cite{Song:2020score, Ho:2020denoising}.

The VP process is defined by the stochastic differential equation
\begin{equation}
    dX_t = -\gamma(t) X_t \, dt + \sqrt{2\gamma(t)}\, dW_t\,,
\end{equation}
where $\gamma(t) > 0$.
The drift term controls the decay of the signal over time. The solution can be written as
\begin{equation}
    X_t
    =
    e^{-\int_0^t \gamma(s)\, ds} X_0
    +
    \sqrt{1 - e^{-2\int_0^t \gamma(s)\, ds}}\, \eta,
\end{equation}
with $\eta \sim \mathcal{N}(0,I)$ independent of $X_0$.

A related construction is the sub-VP process, which keeps the same drift but modifies the noise strength:
\begin{equation}
    dX_t
    =
    -\gamma(t) X_t \, dt
    +
    \sqrt{2\gamma(t)\left(1 - e^{-4\int_0^t \gamma(s)\, ds}\right)}\, dW_t\, .
\end{equation}
Its solution has the form
\begin{equation}
    X_t
    =
    e^{-\int_0^t \gamma(s)\, ds} X_0
    +
    \left(1 - e^{-2\int_0^t \gamma(s)\, ds}\right) \eta\,.
\end{equation}

We also consider a closely related variant,
\begin{equation}
    dX_t
    =
    -\gamma(t) X_t \, dt
    +
    \sqrt{2\gamma(t)\left(1 - e^{-\int_0^t \gamma(s)\, ds}\right)}\, dW_t.
    \label{eq:alternative-sub-VP}
\end{equation}
The corresponding solution is
\begin{equation}
    X_t
    =
    e^{-\int_0^t \gamma(s)\, ds} X_0
    +
    \left(1 - e^{-\int_0^t \gamma(s)\, ds}\right) \eta.
\end{equation}

All three processes share the same effective signal decay factor.
Choosing
\begin{equation}
    \gamma(t) = \frac{1}{1-t}\,,
\end{equation}
leads to a linear decay of the signal term since
\begin{equation}
    e^{-\int_0^t \gamma(s)\, ds} = 1 - t\,.
\end{equation}
In particular, for the last process one obtains
\begin{equation}
    X_t = (1 - t) X_0 + t X_1\,,
    \label{eq:linear-interpolation}
\end{equation}
where $X_1 = \eta$.
This corresponds to a linear interpolation between data and noise, as commonly used in flow-matching and optimal transport formulations for generative modeling~\cite{Liu:2022flow, Lipman:2022flow}.

Overall, this shows that linear decay of signal can be obtained by a specific choice of drift term.
In the Lie-group setting studied in this work, a similar linear decay of observables arises instead from the deterministic drift term generated by the stochastic evolution itself.

\section{Summary and concluding remarks}
\label{sec:conclusion}

We studied how the choice of noise schedule controls the evolution of observables in diffusion processes for lattice gauge theory.
A key observation is that the stochastic evolution of link variables induces a deterministic drift term through It\^o calculus. This reduces the dynamics of gauge-invariant observables to simple evolution equations whose coefficients depend on the noise schedule.

With the noise schedule in eq.~\eqref{eq:noise:schedule} and an appropriate normalization, the Wilson gauge action exhibits linear decay in diffusion time. The same holds for Wilson loops of different sizes after rescaling the noise amplitude. In contrast, Euclidean diffusion models require an explicitly designed drift term to obtain similar linear behavior, as in eq.~\eqref{eq:linear-interpolation}.

In the study of $\mathrm{SU}(3)$ gauge theory in ref.~\cite{Komijani:2026lan}, the normalization of the schedule was chosen empirically to obtain approximately linear evolution of the average plaquette. That normalization is close to the analytical normalization derived here, $\sigma_0 = 1/\sqrt{2C_F} = \sqrt{3}/4$ for $N=3$.

The choice of noise schedule also affects the time dependence of the score function and, consequently, the numerical integration of the reverse diffusion process. In particular, schedules that induce nearly linear dynamics are less affected by discretization errors and can be integrated accurately with a small number of steps. In many experiments of ref.~\cite{Komijani:2026lan}, this allows the reverse process to be solved with very few steps.
This suggests that noise scheduling is a simple and effective handle for improving the efficiency of diffusion-based sampling.

\acknowledgments

The author thanks Andreas Ipp, Marina K. Marinkovic, Thomas R. Ranner, Lara Turgut, and Octavio Vega for useful discussions about the noise schedule.

\bibliographystyle{apsrev4-1}
\bibliography{references.bib} 

\end{document}